\documentclass[aps,prc,preprint,groupedaddress]{revtex4}

\newcommand{\beq}{\begin{equation}}
\newcommand{\eeq}{\end{equation}}
\newcommand{\beqa}{\begin{eqnarray}}
\newcommand{\eeqa}{\end{eqnarray}}

\usepackage{epsfig}

\begin{document}

 \title{The small axial charge of the $N(1535)$ resonance}

\author{C. S. An}
\email[]{ancs@ihep.ac.cn} \affiliation{Institute of High Energy
Physics, CAS, P.O.Box 918, Beijing 10049, China}

\author{D. O. Riska}
\email[]{riska@pcu.helsinki.fi}
\affiliation{Helsinki Institute of Physics, POB 64,
00014 University of Helsinki, Finland}

\thispagestyle{empty}

\date{\today}
\begin{abstract}

There is a natural cancellation between the contributions of the
$qqq$ and $qqqq\bar q$ components to the axial charge of the
$N(1535)$ resonance. While the probability of the former is larger
than that of the latter, its coefficient in the axial charge
expression is exceptionally small. The magnitude of two of the
corresponding coefficients of the $qqqq\bar q$ components are in
contrast large and have the opposite sign. This result provides a
phenomenological illustration of the recent unquenched lattice
calculation result that the axial charge of the $N(1535)$ resonance
is very small, if not vanishing \cite{takah}. The result sets an
upper limit on the magnitude of the probability of $qqqq\bar q$
components as well.

\end{abstract}


\maketitle

A number of phenomenological failures of the constituent quark model
for the baryons may be repaired by extending the model space beyond
that of the basic three quark configurations $qqq$
\cite{Li1,Li2,Li3}. The question of key interest is then that of the
relative magnitude of the sea-quark configurations, and in
particular of the most obvious $qqqq\bar q$ configurations. For most
electromagnetic and strong decay observables, this is difficult to
estimate, because of the very strong contribution from the
transition matrix elements between the $qqq$ and $qqqq\bar q$
components \cite{An}. The axial current operator of the baryon
resonances is an exception, as for this the transition matrix
elements are suppressed - i.e. they involve the small components of
the spinors - with respect to the diagonal matrix elements, so that
the axial charges, to a good approximation, may be expressed as a
sum of the diagonal matrix elements of all possible configurations,
which takes the form of numerical coefficients $A_n$ times the
corresponding probabilities $P_n$:
\begin{equation}
 g_A^* \simeq \sum_n A_n P_n\, .
 \label{gA}
\end{equation}

The (diagonal) axial charges of baryon resonances are however not
accessible experimentally. It is in this regard that the recent
result, obtained numerically by an unquenched QCD lattice
calculation, that the axial charge of the $N(1535)$ actually may
vanish in the two-flavor case, is so interesting \cite{takah}. As
that result appears to be insensitive to the quark mass (the
magnitude of the value extrapolated to 0 is less than 0.2), it may
be taken as a substitute for an experimental value. While the
statistical error margins of the calculated values of the axial
charge of the $N(1535)$ are not yet sufficiently narrow to exclude
the small value -1/9 given by the conventional constituent quark
model with only $qqq$ configurations \cite{gloz}, it is interesting
to explore the phenomenological consequences of a vanishing axial
charge.

If the axial charge of the $N(1535)$ vanishes, it implies that the
sea-quark configurations shall have to cancel the (small)
contribution of the $qqq$ configuration. This makes it possible to
put constraints on the sea-quark configurations in the $N(1535)$. To
illustrate this possibility we consider the contributions from all
the $qqqq\bar q$ components, which may exist in the $N(1535)$. These
have been enumerated in ref. \cite{Helminen}. As all 5 constituents
in a $qqqq\bar q$ configuration in the negative parity $N(1535)$ may
be in the ground state, the orbital state of the 4 quarks may be
assumed to be completely symmetric. Then either the spin-flavor
state has to have the mixed flavor-spin symmetry $[31]_{FS}$ or
alternatively the color-spin state has to have one of the mixed
flavor symmetries $[31]_{CS}$, $[22]_{CS}$ or $[211]_{[CS]}$. There
are 5 different $qqqq\bar q$ configurations in the $N(1535)$ that
have an appropriate symmetry structure and spin and isospin $1/2$.
These are listed in Table \ref{conf}.

\begin{table}
\caption{The $qqqq\bar q$ configurations in the $N(1535)$
and the corresponding axial charge coefficient $A_n$ (\ref{gA}).
\label{conf}}
\begin{tabular}{|c|c|c|c|c|c|}
\hline
configuration & flavor-spin&$C_{FS}$&color-spin&$C_{CS}$&$A_n$
\\
\hline
1& $ [31]_{FS}[211]_F [22]_S $ & $-16$
& $ [31]_{CS}[211]_C [22]_S $& $-16$&  $0$ \\
2& $ [31]_{FS}[211]_F [31]_S $ & $-40/3$
& $ [31]_{CS}[211]_C [31]_S $&$-40/3$&  $+5/6$   \\
3& $ [31]_{FS}[22]_F  [31]_S $ & $-28/3$
& $ [22]_{CS}[211]_C [31]_S $& $-16/3$&  $-1/9$   \\
4& $ [31]_{FS}[31]_F  [22]_S $ &$-8$
& $ [211]_{CS}[211]_C [22]_S $ & $0$&  $-4/15$ \\
5& $ [31]_{FS}[31]_F  [31]_S $ & $-16/3$
& $ [211]_{CS}[211]_C [31]_S $& $+8/3$&$+17/18$\\
\hline
\end{tabular}
\end{table}

The numbering of these configurations are in order of increasing
energy, if the hyperfine interaction between the quarks is assumed
to depend either on flavor and spin or on color and spin. In the
table the matrix elements of the schematic hyperfine splitting
operator
\begin{equation}
C_{kS}=-\sum_{i,j}\vec\lambda_i\cdot \vec\lambda_j
\vec\sigma_i\cdot\vec\sigma_j \label{hf}
\end{equation}
are listed for both the cases where the operators $\vec\lambda$
represent either the generators of the color $SU(3)$ (k=$C$) or the
flavor $SU(3)$ group (k=$F$), respectively. (Here the spatial
structure of the interaction has been neglected, as all the
constituents are in the same orbital ground state). Note that
because of their mixed flavor symmetry $[211]_F$ both the
configurations (1) and (2) in Table \ref{conf} have to contain a
strange quark-antiquark pair. This is as expected on the basis of
the observed large $N\eta$ decay branch of the $N(1535)$.

Because the states 1 in Table \ref{conf} has zero total spin, and
the antiquark is a strange quark, it contributes no matrix element
to the axial charge operator $\sum_i \sigma_z(i)\tau_z(i)$
(calculated here as the matrix element of the third component of the
axial vector current). In the table the matrix element of the axial
charge of these configurations, combined with the wave function of
the antiquark are also listed. The general expression in the
flavor-spin coupling scheme for these 5 quark wave functions is :
\begin{eqnarray}
&\psi_{t,s}^{(i)} =\sum_{a,b,c}\sum_{Y,y,T_z,t_z}\sum_{S_z,s_z}
C^{[1^4]}_{[31]_a[211]_a} C^{[31]_a}_{[F^{(i)}]_b [S^{(i)}]_c}
[F^{(i)}]_{b,Y,T_z} [S^{(i)}]_{c,S_z}
[211;C]_a\nonumber\\
&(Y,T,T_z,y,\bar t,t_z|1,1/2,t)
(S,S_z,1/2,s_z|1/2,s)\bar\chi_{y,t_z}\bar\xi_{s_z}\varphi_{[5]}\, .
\label{wfc}
\end{eqnarray}
Here $i$ is the number of the $qqqq\bar q$ configuration in Table
\ref{conf}, $\bar\chi_{y,t_z}$ and $\bar\xi_{s_z}$ represent the
isospinor and the spinor of the antiquark respectively, and
$\varphi_{[5]}$ represents the completely symmetrical orbital wave
function. The first summation involves The symbols
$C^{[.]}_{[..][...]}$, which are $S_4$ Clebsch-Gordan coefficients
for the indicated color ($[211]$), flavor-spin ($[31]$) and flavor
($[F]$) and spin ($[S]$) wave functions of the $qqqq$ system. The
second summation runs over the flavor indices in the $SU(3)$
Clebsch-Gordan coefficient (with 9 symbols) and the third over the
spin indices in the standard $SU(2)$ Clebsch-Gordan coefficient. In
the case of the spin configuration $[22]$ the total spin of the
$qqqq$ system vanishes, so that $S=S_z=0$. These wave functions are
given in explicit form in Ref. \cite{An2}.

With the results in Table \ref{conf}, the explicit expression for the
axial charge of the $N(1535)$ takes the form
\begin{equation}
g_A (N(1535))= -{1\over 9}P_3 + {5\over 6} P_5^{(2)}-{1\over 9}
P_5^{(3)}-{4\over 15}P_5^{(4)} +{17\over 18}P_5^{(5)}\, .
\label{gax}
\end{equation}
Here $P_3$ is the probability for the conventional $qqq$
configuration, while $P_5^{(i)}$ represents the probabilities of the
$qqqq\bar q$ configurations in Table \ref{conf}. Note that the
energetically most favorable $qqqq\bar q$ configuration (1) does not
contribute to the axial charge at all.

The fact that the two $qqqq\bar q$ contributions in (\ref{gax}),
which are positive, have large coefficients $\sim 1$, while the
coefficient of the $qqq$ contribution is small and negative ($-1/9$)
immediately suggests the possibility for a considerable cancellation
between the $qqq$ valence and the $qqqq\bar q$ sea-quark
contributions, as the probability of the latter is likely to be
considerably smaller than that of the former. If only the first two
terms in the expression ({\ref{gax}}) are taken into account
$g_A(N(1535))$ would vanish if $P_5^{(2)} = 2/15 P_{qqq}$, which may
be a fairly reasonable assumption. The last two remaining $qqqq\bar
q$ configurations are in expected to have very small probability, as
they are energetically unfavorable (Table \ref{conf}).

In ref. \cite{An2} it was in fact found that the quark model
prediction for the helicity amplitude $A_{1/2}$ for
$N(1535)\rightarrow N\gamma$ could be brought qualitatively into
line with the empirical values if $P_{qqq}\simeq 0.55$ and
$P_5^{(1)}\simeq 0.45$. Since the $qqqq\bar q$ configurations (1)
and (2) in Table \ref{conf} are similar in that both involve an
$s\bar s$ pair, but the latter is energetically disfavored by the
matrix elements of the hyperfine interaction (\ref{hf}), the
helicity amplitude should be similar if the probability $P_5^{(2)}$
for the configuration (2) in Table \ref{conf}, which has a large
axial charge coefficient (\ref{gax}), falls in the range (0.25-0.3)$
P_5^{(1)}$. With these numbers $g_A(N(1535))$ comes out to lie in
the range 0.03-0.06. If on the other hand one considers both the
configurations (2) and (3) in Table \ref{conf} as equally probable:
$P_5^{(2)}=P_5^{(3)}$ and the probabilities of to fall within the
range (0.12-0.15)$ P_5^{(1)}$, the numerical value for
$g_A(N(1535))$ falls in the range -0.02 to -0.05. This shows that
the likely range of values for $g_A(N(1535))$ in the extended quark
model, which includes explicit $qqqq\bar q$ components  ~ -0.05 ..
+0.06, brackets 0. This range would bracket 0 also in the case where
the relative $qqq$ probability where increased to $P_3=0.7$ and
$P_5^{(2)}=0.3$. It does in any case not appear possible to reach
the value 0 for $g_A(N(1535))$, with an overall $qqqq\bar q$
probability that is larger than 0.45.

The conclusion is therefore that the very small or possibly
vanishing axial charge of the $N(1535)$ already at the present level
of accuracy constrains the magnitude of the probability of the
sea-quark components in the $N(1535)$ to be less than 45\%. A more
general observation is that the axial charges of the baryon
resonances may be useful for setting limits on the probabilities of
their sea-quark configurations.

\begin{acknowledgments}

D. O. Riska thanks Dr. L. Ya. Glozman for drawing attention to the
axial charge of the $N(1535)$. C. S. An acknowledges the hospitality
of the Helsinki Institute of Physics during the course of this work.
This work is partly supported by the National Natural Science
Foundation of China under grants Nos. 10435080, 10521003, and by the
Chinese Academy of Sciences under project No.~KJCX3-SYW-N2.

\end{acknowledgments}

\end{document}